\documentclass[final,5p,times,twocolumn,draft]{elsarticle}

\usepackage{amssymb}
\usepackage{amsmath}
\usepackage{bm}
\usepackage{mathrsfs}
\usepackage{textcomp}
\usepackage{hyperref}
\usepackage{lineno}
 \usepackage{placeins}
\usepackage{comment}
\usepackage{natbib}
\biboptions{sort&compress}
\usepackage{setspace}
\usepackage{algpseudocode}

\renewcommand{\eqref}[1]{\textup{{\normalfont Eq.~(\ref{#1}}\normalfont)}}

\usepackage{ulem}
\usepackage{xcolor}

\DeclareRobustCommand{\Erase}{\bgroup\markoverwith{\textcolor{red}{\rule[.5ex]{2pt}{0.4pt}}}\ULon}

\begin{document}

\begin{frontmatter}

\title{Self-locking and Stability of the Bowline Knot}

\author[a,b]{Bastien F.G. Aymon}
\author[a,c]{Fani Derveni}
\author[a,d]{Michael Gomez}
\author[e,f]{Jérôme Crassous}
\author[a]{Pedro M. Reis\corref{cor1}}

\address[a]{Flexible Structures Laboratory, Institute of Mechanical Engineering, École Polytechnique Fédérale de Lausanne (EPFL), 1015 Lausanne, Switzerland}

\address[b]{Department of Mechanical Engineering, Massachusetts Institute of Technology, Cambridge, MA 02138, USA}

\address[c]{Scalable MetaStructures Laboratory, School of Civil and Environmental Engineering, Cornell University, Ithaca, NY 14853, USA}

\address[d]{Department of Engineering, King's College London, Strand, London WC2R 2LS, United Kingdom}

\address[e]{University of Rennes, CNRS, IPR (Institut de Physique de Rennes) - UMR 6251, F-35000 Rennes, France}

\address[f] {PMMH, CNRS, ESPCI Paris, Université PSL, Sorbonne Université, Université Paris Cité, F-75005 Paris, France}

\cortext[cor1]{Corresponding author. E-mail address: pedro.reis@epfl.ch (P.M. Reis)}

\begin{abstract}
We investigate the self-locking of the bowline knot through numerical simulations, experiments, and theoretical analysis. Specifically, we perform two complementary types of simulations using the 3D finite-element method (FEM) and a reduced-order model based on the discrete-element method (DEM). For the FEM simulations, we develop a novel mapping technique that automatically transforms the centerline of the rod into the required knot topology prior to loading. In parallel, we conduct experiments using a nearly inextensible elastic rod tied into a bowline around a rigid cylinder. One end of the rod is pulled to load the knot while the other is left free. The measured force-displacement response serves to validate both the FEM and DEM simulations. Leveraging these validated computational frameworks, we analyze the internal tension profile along the rod's centerline, revealing that a sharp drop in tension concentrates around a strategic locking region, whose geometry resembles that observed in other knot types.  By considering the coupling of tension, bending, and friction, we formulate a theoretical model inspired by the classic capstan problem to predict the stability conditions of the bowline, finding excellent agreement with our FEM and DEM simulations. Our methodology and findings offer new tools and insights for future studies on the performance and reliability of other complex knots.
\end{abstract}

\end{frontmatter}

\section{Introduction}
\label{Introduction}

\begin{figure*}[h!]
    \centering
    \includegraphics[width=0.6\textwidth]{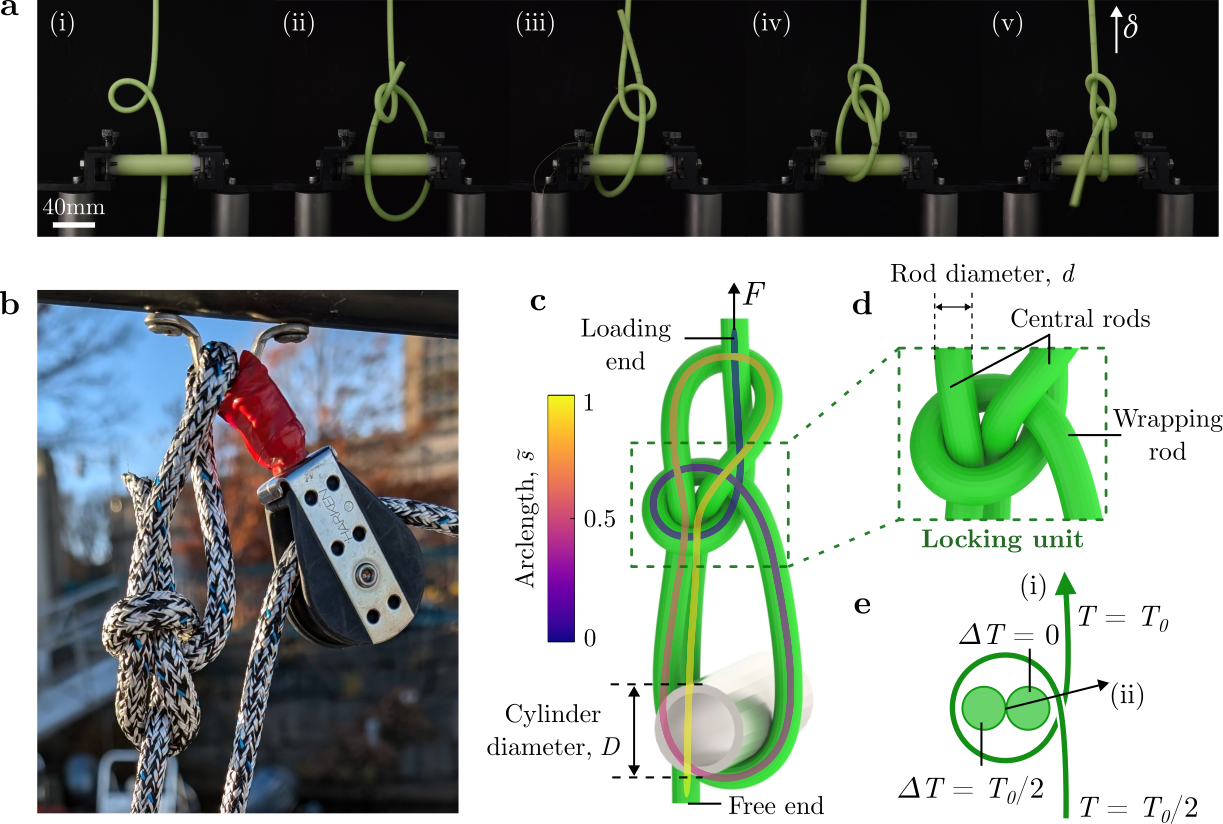}
    \caption{The bowline knot. \textbf{a} Experimental knot tying sequence. Following step (v), a displacement $\delta$, at constant velocity, $v$, is imposed at the top end of the knot. \textbf{b} Example of a bowline knot in a practical setting: a pulley system on a boat's sail. \textbf{c} Geometry and primary features of the bowline. \textbf{d} Close-up and \textbf{e} schematic of the locking unit, with tension values marked at specific locations. The bowline can unravel via either the sliding of (i) the wrapping rod or (ii) the central rods.}
    \label{fig:fig1}
\end{figure*}

The bowline knot, often nicknamed the `\textit{king of the knots},' is so fundamental to maritime operations that Clifford Ashley, in his seminal book~\cite{Ashley1944}, states: `\textit{A sailor seldom uses another loop knot aboard ship}.' The bowline owes its popularity to its robustness and an intuitive tying method (Fig.~\ref{fig:fig1}\textbf{a}), passed down through generations via the familiar mnemonic: `\textit{The rabbit comes out of the hole, runs around the tree, and goes back in the hole}.' A bowline may be readily attached to a mooring post, cringles (rigid rings) in a sail, or even used as an emergency rescue harness around the waist of a sailor. The bowline creates a secure, non-slip loop at the end of a rope (Fig.~\ref{fig:fig1}\textbf{b}), forming a \textit{self-locking structure} that can be easily untied even after bearing heavy loads. This self-locking mechanism is not unique to the bowline and appears in several other knots employed in sailing and climbing, including the sheepshank, the double fisherman's knot, the alpine butterfly, and perfection loops~\cite{Ashley1944}.

While knot theory is a well-established field in mathematics, particularly in topology ~\cite{Rolfsen_2003,Adams_2010} and geometry \cite{Gonzalez_Maddocks_1999,Carlen_Laurie_Maddocks_Smutny_2005}, the mechanics of physical knots remains largely unexplored. Recent mechanics-based studies of some classical knots (\textit{e.g.}, the overhand ~\cite{Audoly_Clauvelin_Neukirch_2007,Jawed_Dieleman_Audoly_Reis_2015,Baek_Johanns_Sano_Grandgeorge_Reis_2020}, trefoil~\cite{Johanns_Grandgeorge_Baek_Sano_Maddocks_Reis_2021}, fisherman's ~\cite{Tong_Ibrahim_Khalil_Justin_Silva_Wang_Khoda_Khalid_Jawed_2023}, clove hitch~\cite{Sano_Johanns_Grandgeorge_Baek_Reis_2022},  surgical~\cite{Johanns_Baek_Grandgeorge_Guerid_Chester_Reis_2023} and stopper~\cite{Johanns_Reis_2024} knots) have revealed how the interplay between topology, geometry, elasticity, and friction underlies the remarkable performance of these seemingly simple structures. Building on these recent advances, we study the \textit{bowline}; a knot that, despite centuries of use, lacks thorough mechanical characterization and structural analysis.

Here, we combine precision experiments, simulations, and theoretical analysis to investigate the mechanics of the bowline knot, focusing on its self-locking behavior. First, we introduce a novel mapping technique to efficiently create initial knotted configurations for finite element method (FEM) simulations. Previous approaches \cite{Baek_Johanns_Sano_Grandgeorge_Reis_2020,Johanns_Grandgeorge_Baek_Sano_Maddocks_Reis_2021,Johanns_Baek_Grandgeorge_Guerid_Chester_Reis_2023,Johanns_Reis_2024} required time-consuming and cumbersome step-by-step knot construction. By contrast, our approach automatically transforms centerline geometry into the desired configuration through optimized `temporary' displacements, significantly reducing computational time. We validate this framework through experiments on a nearly inextensible elastomeric rod and compare these results to discrete element method (DEM) simulations for complementary insights. Examining the tension profile along the rod centerline reveals that most force drop localizes around a strategic locking region. This motivates an analytical stability criterion based on the classic capstan problem, considering the interplay between frictional, axial, and bending forces. Finally, this criterion is tested using DEM and FEM results. Our framework provides a foundation for quantitatively investigating the stability of more complex physical knots.

\section{Definition of the problem}
\label{sec:problem_definition}

We consider the bowline knot (see schematic in Fig.~\ref{fig:fig1}\textbf{c}) obtained by tying a rod of length $L$ and diameter $d{=}2r$ around a rigid cylinder of diameter $D$. First, a small loop, often referred to as the `\textit{rabbit hole}', is formed in the portion of the rod above the cylinder (Fig.~\ref{fig:fig1}\textbf{a},i). Next, the rod's free end (the `\textit{rabbit}') is passed under the cylinder and then upward through this loop (Fig.~\ref{fig:fig1}\textbf{a},ii). The free end is then routed around the upper segment of the rod (the `\textit{tree}'; Fig.~\ref{fig:fig1}\textbf{a},iii) before being guided back down through the loop again (Fig.~\ref{fig:fig1}\textbf{a},iv). Finally, the knot is tightened by pulling the upper segment of the rod (Fig.~\ref{fig:fig1}\textbf{a},v). An upward displacement, $\delta$, is applied quasi-statically at the loading (upper) end of the rod; the corresponding tightening force is $F$. The lower end of the rod is left free. The normalized arclength along the centerline, $s/L{=}\tilde{s} \in [0,1]$, is chosen to start at the loading end and finish at the free end. 

The key feature of the bowline knot we seek to investigate is its \textit{locking unit}: the region where a `wrapping rod' segment loops around two central rod segments (Fig.~\ref{fig:fig1}\textbf{d}). The importance of this locking unit is evidenced by the fact that a bowline under tension will remain stable if the large loop formed by the central rods is cut, provided that the locking unit is undisturbed (see Supplementary Video 1). We aim to understand how this locking unit contributes to the bowline's self-locking ability, overall stability, and, more broadly, to its widespread popularity and reliability. We also assess the capacity of reduced-order models --- including DEM simulations of elastic rods and simpler analytical frameworks --- to accurately capture the complex mechanics of tight physical knots such as the bowline.

\section{Methods - Numerical simulations: FEM and DEM}
\label{sec:method_FEM}

Several analytical models for elastic knots have been developed in the literature~\cite{Audoly_Clauvelin_Neukirch_2007,Jawed_Dieleman_Audoly_Reis_2015}, though they tend to focus on \textit{loose} configurations. By contrast, \textit{tight} physical knots, where bending curvatures are comparable to the rod diameter, are more relevant to applications but call for complex simulations. Our group recently introduced a general strategy to simulate tight knots in 3D FEM~\cite{Baek_Johanns_Sano_Grandgeorge_Reis_2020}. This framework was subsequently used to simulate a variety of tight elastic knots~\cite{Baek_Johanns_Sano_Grandgeorge_Reis_2020,Johanns_Grandgeorge_Baek_Sano_Maddocks_Reis_2021,Johanns_Reis_2024}, and extended to incorporate elasto-plastic constitutive behavior, which is relevant for knots in polymeric surgical filaments~\cite{Johanns_Baek_Grandgeorge_Guerid_Chester_Reis_2023}.

State-of-the-art 3D FEM simulations of knots are time-consuming and cumbersome. The first important step in setting up an FEM model for tight knots is obtaining the appropriate topology of the knot in a loose configuration prior to loading. Existing approaches for this procedure involve a series of carefully chosen tying sub-steps~\cite{Baek_Johanns_Sano_Grandgeorge_Reis_2020} --- including repeatedly pinning or clamping nodes of interest and loading others --- that are highly specific to the chosen knot. Extensive trial and error is required to ensure that the simulation converges in a reasonable number of steps and the final configuration looks as desired. These computational challenges are especially problematic for the bowline knot, since its topology is considerably more intricate than previously investigated knots (\textit{e.g.}, overhand, trefoil, clove hitch) with more symmetric, compact structures. 

Reduced-order models based on the Discrete Elastic Rod (DER) method~\cite{Bergou_Wardetzky_Robinson_Audoly_Grinspun_2008, jawed_coiling_rigid_substrates} are another emerging tool to simulate the mechanics of elastic knots~\cite{Tong_Ibrahim_Khalil_Justin_Silva_Wang_Khoda_Khalid_Jawed_2023}. These numerical frameworks are significantly more computationally efficient than 3D FEM, though they neglect cross-section deformations, which are expected to be significant for tight knots where large contact pressures occur. Indeed, Grandgeorge \textit{et al.}~\cite{Grandgeorge_Baek_Singh_Johanns_Sano_Flynn_Maddocks_Reis_2021} have shown that cross-sectional deformation can occur in the canonical case of two elastic filaments in tight orthogonal contact. Still, it is unclear to what extent cross-sectional deformations influence the large-scale mechanical response of tight knots.

In light of the above, we introduce a novel \textit{mapping technique} that automatically ties an initially straight rod into any desired knot shape in 3D FEM, orders of magnitude faster than existing strategies. We describe the implementation of this mapping technique into a commercial FEM software. Finally, we also employ a DEM simulation framework developed recently for frictional fibers~\cite{Crassous_2023}, which assumes rigid rod cross-sections. 

\subsection{A centerline-displacement mapping for elastic knots}\label{sec:mappingTechniqueMain}

In this section, we outline the mapping technique we used to set up the initial configuration of the bowline knot for FEM simulations; additional details are provided in \ref{app:directMapping}. 

\begin{figure}[h!]
    \centering
    \includegraphics[width=\columnwidth]{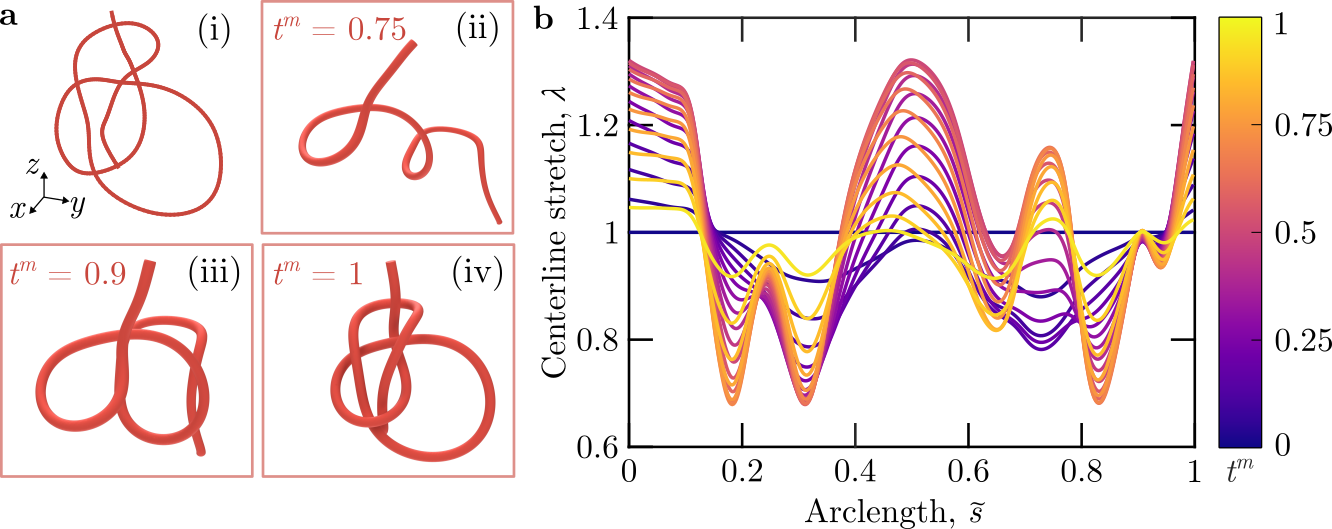}
    \caption{Mapping technique for initial configuration of the FEM simulations. \textbf{a} (i) `Raw' curve (before Laplacian smoothing is applied) of a loose bowline knot, hand-drawn using Blender. (ii)-(iv) Snapshots of the rod taken from the FEM model along different instances of the mapping iteration time, $t^m$. Contact is momentarily turned off during this phase. \textbf{b} Centerline stretch, $\lambda$, as a function of the normalized rod arclength, $\tilde{s}$. Curves are color-coded by the mapping time, $t^{m}$.  The centerline stretch is measured relative to the straight, undeformed configuration; $\lambda=1$ corresponds to an inextensible mapping of the straight rod.}
    \label{fig:fig2}
\end{figure}

Our method utilizes a displacement mapping function designed to minimize stretching along the centerline during the mapping phase, facilitating convergence in the FEM solver. First, we hand-draw a space curve with the desired topology of the bowline using a non-uniform rational b-spline (NURBS) curve in Blender~\cite{blender}, as shown in Fig. \ref{fig:fig2}\textbf{a}(i), rather than relying on mathematical parameterizations of ideal knots~\cite{Adams_2010} or experimental measurements (\textit{e.g.}, using X-ray tomography), which can be limited or cumbersome. Subsequently, we discretize the resulting curve into vertices and define a uniform mesh size, $h$, along the centerline, whose integral yields the rod length $L$. Local curvature artifacts arising from the hand-drawn data are mitigated by applying uniform Laplacian smoothing iteratively to the discretized centerline coordinates; see \ref{app:directMapping} for details. The resulting smoothed coordinates are imported into MATLAB and scaled uniformly to match the rod length ($L{=}600\,$mm) used in the experiments (Section~\ref{sec:rod_fab}). Hereafter, we denote these final scaled and smoothed coordinates as $\mathbf{x}_i$.

Given the specified mesh size $h$, we define the displacement between the initial and knotted configurations at each vertex as $\mathbf{u}_i=\mathbf{x}_i-\mathbf{X}_i$, where $\mathbf{X}_i=ih\mathbf{e}_z$ corresponds to the coordinates in the straight, undeformed rod. Direct application of this displacement field in FEM typically leads to extreme intermediate deformations and convergence issues (see Fig.~\ref{fig:SuppA}). To circumvent these issues, we introduce \textit{temporary displacements}, $\mathbf{v}_i$, which are optimized to minimize intermediate stretches. The displacement mapping function used for this purpose is defined in~\eqref{eq:mapping}. We determine the optimal polynomial coefficients of the temporary displacements via an iterative optimization process, minimizing the maximum centerline stretch, $\Lambda_{\text{max}}$, until a desired threshold ($|\Lambda_{\text{max}} - 1| {<} 0.3$) is reached; see \eqref{eq:param_opti} and \ref{app:directMapping} for details on this optimization procedure.

\subsection{Finite-element simulations\label{subsec:methods_FEM}}

We perform finite element simulations using Abaqus 6.23 (Simulia;~2023). The displacement mapping function described above (cf.~\eqref{eq:mapping}) is applied to all centerline nodes of an initially straight rod via a displacement boundary condition in Abaqus/Standard (Fig.~\ref{fig:fig2}\textbf{a}). We use a custom \texttt{UDISP} user subroutine (see~\ref{app:directMapping}) to overcome Abaqus' limitation in handling time- and node-dependent displacement boundary conditions. Contact is disabled during this initial step. As shown in Fig.~\ref{fig:fig2}\textbf{b}, stretching remains low throughout the mapping because of the temporary displacements, guaranteeing convergence of the FEM solver. All subsequent steps are run in Abaqus using a nonlinear dynamic-implicit analysis. 

At the end of the mapping phase, contact is activated using normal penalty forces and a Coulomb friction model. The loose knot is then relaxed with both ends fixed until equilibrium is reached. A controlled force or displacement is finally applied at its loading end, depending on the scenario under consideration. In both cases, loads are applied sufficiently slowly (typical velocity $v/d \approx 0.67\,\mathrm{s}^{-1}$) to ensure negligible dynamic effects. For experimental validation (see Section~\ref{sec:mechanical_test}), self-weight is included using $g{=}9.81\,\mathrm{m}\,\mathrm{s}^{-2}$ for gravitational acceleration.

To simulate the combined Nitinol-elastomer rods used in our experiments (see Section~\ref{sec:rod_fab}), the rod centerline is meshed using T3D2 linear truss elements ($h/L{=}1.67\times10^{-3}$) while the elastomeric bulk of the rod uses C3D8H linear brick elements (44 elements per cross-section), as in previous studies~\cite{Baek_Johanns_Sano_Grandgeorge_Reis_2020,Johanns_Grandgeorge_Baek_Sano_Maddocks_Reis_2021,Johanns_Reis_2024}. The material properties are chosen to match the experiments (see Section~\ref{sec:rod_fab}). The addition of Nitinol to the centerline of the experimental elastomeric rod results in a $14\%$ increase in bending stiffness compared to the monolithic elastomeric rod (see discussion in Section~\ref{sec:rod_fab}); the Young's modulus of VPS in the simulations is increased accordingly. The horizontal rigid cylinder around which the knot is tied is modeled as an analytical rigid shell with diameter $D {=} 20\,\mathrm{mm}$, as in experiments.

The typical FEM computational time to obtain the initial loose knot topology on a workstation using 15 cores (Intel Xeon E5-2670 2.30 GHz) is 14 minutes, which is a significant improvement compared to the 60 hours reported in a previous study for the tying and loading of an overhand knot~\cite{Baek_Johanns_Sano_Grandgeorge_Reis_2020}.

\subsection{Discrete-element simulations}

In parallel with the FEM simulations described above, we conduct Discrete Element Method (DEM) simulations using the recently introduced numerical framework detailed in Ref.~\cite{Crassous_2023}. Briefly, the rod is modeled as a set of cylinders of constant circular sections (\textit{i.e.}, no cross-sectional deformations). The elastic stretching, bending, and twisting forces acting on the rod are computed using a Discrete Elastic Rod (DER) model. In particular, the curvature and torsion of the discrete rod are derived from the change in orientation of the local material frames attached to the cylinder axis. The rod-rod or rod-cylinder contacts are modeled using a Cundall-Strack model~\cite{cundall.1979}; at each contact point, the normal contact force 
    ${\bf f}_n{=}[k_n \delta_p - \lambda_n \dot{\delta}_p] {\bf n}$,
where $\delta_p$ is the cylinder interpenetration, $\dot{\delta}_p$ its time derivative,  $k_n$ the normal contact stiffness,  $\lambda_n$ a damping coefficient, and ${\bf n}$ the common normal to the cylinders at the contact point. The corresponding tangential force between cylinders is computed as ${\bf f}_t{=}k_t {\bf u}_t$, with the tangential contact stiffness $k_t$, and the relative tangential displacement ${\bf u}_t$. This displacement is eventually limited to ensure the friction condition $\vert {\bf f}_t \vert \le \mu_{\text{rr}} \vert {\bf f}_n \vert$, where $\mu_{\text{rr}}$ is the rod-rod friction coefficient.

In all DEM simulations conducted, the bowline is modeled as a set of $N_{c}{=}300$ cylinders of simulation length $l^*{=}1$ and radius $r^* {=} 1.5$, with a physical length scale $l_0{=}2$\,mm (the `0' subscript denotes a physical scale variable). Therefore, the total length of the rod is $L{=}N_{c} l^* l_0{=}600$\,mm, and its diameter is $d{=}2r^*l_0{=}6$\,mm, matching experiments. Taking the axial stiffness of one segment, $k_0$, for the stiffness scale, the physical scales are $f_0{=}k_0\,l_0$ for force and $B_0{=}C_0{=}k_0\,l_0^3$ for the bending and torsional moduli, respectively.

Due to friction, the geometry of the simulated knots can depend on their preparation history for a given set of applied forces. To ensure reproducibility, we initialize the DEM simulations with a configuration (centerline, twist, contact forces) matching that of FEM. We then set $\mu_{\text{rr}}{=}0$ with a frozen centerline and let segments rotate freely to minimize the twisting energy. Subsequently, we set $\mu_{\text{rr}}$ to the desired value and let the knot relax further prior to applying boundary conditions. We note that Section~\ref{sec:stability} involves very soft rods that cannot be simulated using FEM; here, the initial shape is obtained by using a rod configuration composed of a succession of straight segments and circular arcs that replicate the bowline topology. 

The typical DEM computational time to reach equilibrium from an arbitrary initial shape of the rod is typically $30~s$ (on a laptop using an Intel Core Ultra 7 165H), nearly 30 times faster than the FEM simulations.

\section{Methods - Experiments}
\label{sec:rod_fab}

We performed experiments on a soft polymeric rod containing a stiff Nitinol filament at its core. The composite structure is considerably stiffer axially than in bending, qualitatively similar to a rope, while remaining simpler to characterize and model (a rope has an intricate multiscale braided substructure that also involves irreversible deformation due to internal friction of these substructures). The large axial stiffness also allows us to reliably investigate the knot's self-locking ability: without the Nitinol core, the response under tightening would be dominated by axial extension of the unknotted part of the rod. We cast straight, circular rods using vinylpolysiloxane (VPS32, Elite Double 32, Zhermack; Young's modulus $E{=}1.25\,\mathrm{MPa}$, density $\rho {=} 1160\,\mathrm{kg}\,\mathrm{m}^{-3}$) inside a straight steel tube (inner diameter $d{=}2r{=}6\,\mathrm{mm}$, length $L {=} 600\,\mathrm{mm}$), as described previously~\cite{Grandgeorge_Baek_Singh_Johanns_Sano_Flynn_Maddocks_Reis_2021}. Following recent work from our group~\cite{Johanns_Reis_2024}, a thin Nitinol wire (radius $r_{n}{=}0.127\,\mathrm{mm}$, Young's modulus $E{=}75.5\,\mathrm{GPa}$) was suspended along the center of the tube (the mold), with a dead weight ($m{=}200\,\mathrm{g}$) attached at its lower end to ensure that the Nitinol remained straight during casting and curing of the VPS.

The fabricated elastic rods were surface-treated with talcum powder (Millette, Migros) and left to rest for 7 days prior to testing, to stabilize their mechanical properties. This method yields a consistent rod-rod kinetic friction coefficient, measured to be $\mu_{\text{rr}}{=}0.33\pm 0.04$. The rigid cylinder around which the knot is tied (diameter $D {=} 20\,\mathrm{mm}$; see Fig.~\ref{fig:fig1}\textbf{c}) was coated with a thin layer (thickness $t {=} 0.3\,\mathrm{mm}$) of VPS32 and, similarly to the rod, was also surface treated with talcum powder; the measured rod-cylinder kinetic friction coefficient is $\mu_{rc}{=}0.39\pm 0.02$. The frictional behavior of the powder-treated rods and cylinder was systematically characterized following a previously established procedure~\cite{Grandgeorge_Baek_Singh_Johanns_Sano_Flynn_Maddocks_Reis_2021}. 
Even if small, the increase in bending stiffness due to the Nitinol core was still characterized experimentally and taken into account for simulations, as explained in Section~\ref{subsec:methods_FEM}. 
 Using cantilever tests under self-weight, the VPS-Nitinol rods were measured to have a bending stiffness $(EI)_{\text{VPS/Ni}}^{\text{exp}}{=}89.4\pm 8.0\,\mathrm{MPa}\,\mathrm{mm}^4$, whereas bulk VPS rods have $(EI)_{\text{VPS}}^{\text{exp}}{=}78.7\pm 7.5\,\mathrm{MPa}\,\mathrm{mm}^4$.

\section{Mechanical testing of the bowline knot}
\label{sec:mechanical_test}

Bowline knots were tied on a VPS-Nitinol rod around a rigid cylinder  (see Fig.~\ref{fig:fig1}\textbf{a} and Supplementary Video 2). Tensile experiments were performed using a Universal Testing Machine (Instron 2530, 50N load cell). We measured the force $F$ as a function of the displacement $\delta$ imposed at the loading (upper) end of the rod, at a constant pulling velocity of $v{=}4\,\mathrm{mm}\,\mathrm{s}^{-1}$ to ensure quasi-static conditions. The experimentally applied displacement is stopped at $\delta/r{\approx} 33$, as further deformation risks damaging the rod through delamination between the Nitinol core and VPS. The experimental results described in this section will serve to validate the 3D FEM and DEM simulations presented in Section~\ref{sec:method_FEM}.

\begin{figure}[b!]
    \centering
    \includegraphics[width=\columnwidth]{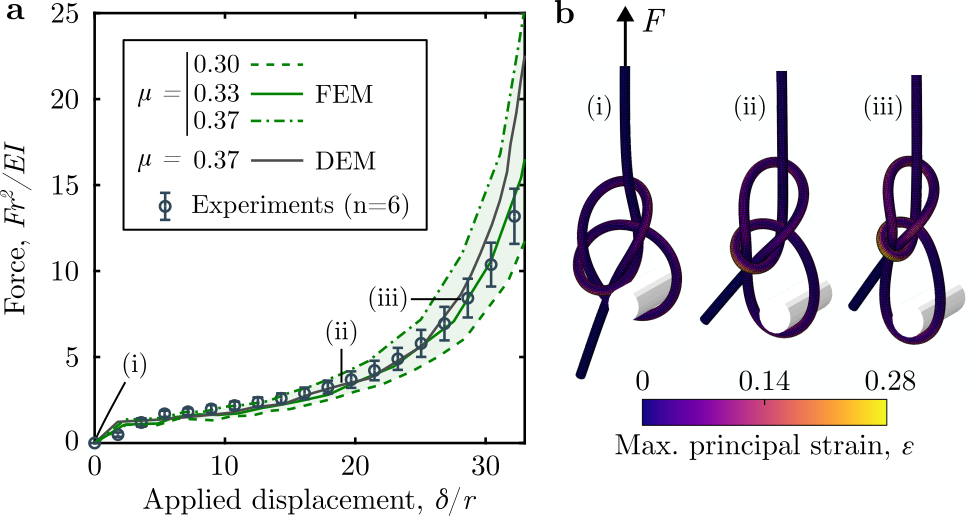}    \caption{Experimental validation of the FEM and DEM frameworks. \textbf{a} Force-displacement curve of the bowline knot obtained from experiments, as well as from FEM and DEM simulations. The experimental data is averaged over six samples, and error bars represent the standard deviation. The shaded area corresponds to lower and upper bounds for the friction coefficient ($\mu_{\text{rr}}{=}0.30$ and $\mu_{\text{rr}}{=}0.37$, respectively) to account for experimental variation. \textbf{b} Snapshots of the knot in FEM at three successive loading steps, (i) to (iii), also represented in \textbf{a}. The colormap corresponds to the maximum principal strain, $\varepsilon$.}
    \label{fig:fig3}
\end{figure}

In Fig.~\ref{fig:fig3}\textbf{a},
we present force-displacement curves for the tightening of the bowline knot obtained from experiments, as well as FEM and DEM simulations. To account for variations in the friction coefficient, we considered $\mu_{\text{rr}}\in\{0.30, 0.33, 0.37\}$ for the FEM and $\mu_{\text{rr}}{=}0.37$ for DEM (the experimental value is $\mu_{\text{rr}}{=}0.33\pm 0.04$; cf. Section~\ref{sec:rod_fab}). Excellent agreement is found between experiments and the two types of simulations,  validating the numerical models, even if some differences appear at higher applied displacements after locking. The force profile is approximately linear for $\delta/r\lesssim 20$, after which it becomes nonlinear, increasing significantly due to self-locking of the knot --- a key feature of the bowline. We remark that the force behavior past the locking onset is highly dependent on the friction coefficient, with the force at the loading end varying by a factor of two between $\mu_{\text{rr}}{=}0.30$ and $\mu_{\text{rr}}{=}0.37$ at $\delta/r{\approx} 33$. This observation evidences the critical role of friction at the onset of locking, which we will investigate further in Section~\ref{sec:stability}. 

Selected snapshots of the FEM tightening procedure along the loading path --- (i) in the loose configuration, (ii) at the onset of nonlinearity, and (iii) at the point of self-locking --- are presented in Fig.~\ref{fig:fig3}\textbf{b}. In these snapshots, we also use a colormap to represent the maximum principal strain, $\varepsilon$, which tends to be largest along the wrapping rod within the locking unit, confirming its role in the knot's self-locking ability.

\section{Stability of the bowline knot}
\label{sec:stability}

Having validated the FEM and DEM numerical frameworks against experiments, we proceed to study the bowline's \textit{stability}. First, we will analyze the tension profile along the bowline. Then, we will leverage this understanding to determine under which material properties (friction coefficient, Young's modulus), geometric parameters (rod radius), and loading conditions the knot remains stable, without unraveling.

\subsection{Tension profile along the centerline of the bowline knot}
\label{subsec:tension}

To investigate how the bowline's geometry makes it so reliable, we quantify the tension profile along its centerline. This quantity is not experimentally available; thus, we must rely solely on the validated FEM and DEM simulations to identify which critical portions of the knot are responsible for reducing tension from a high value (at the loading end) to zero (at the free end), which lies at the core of its self-locking ability. 

\begin{figure}[b!]
    \centering
    \includegraphics[width=0.7\columnwidth]{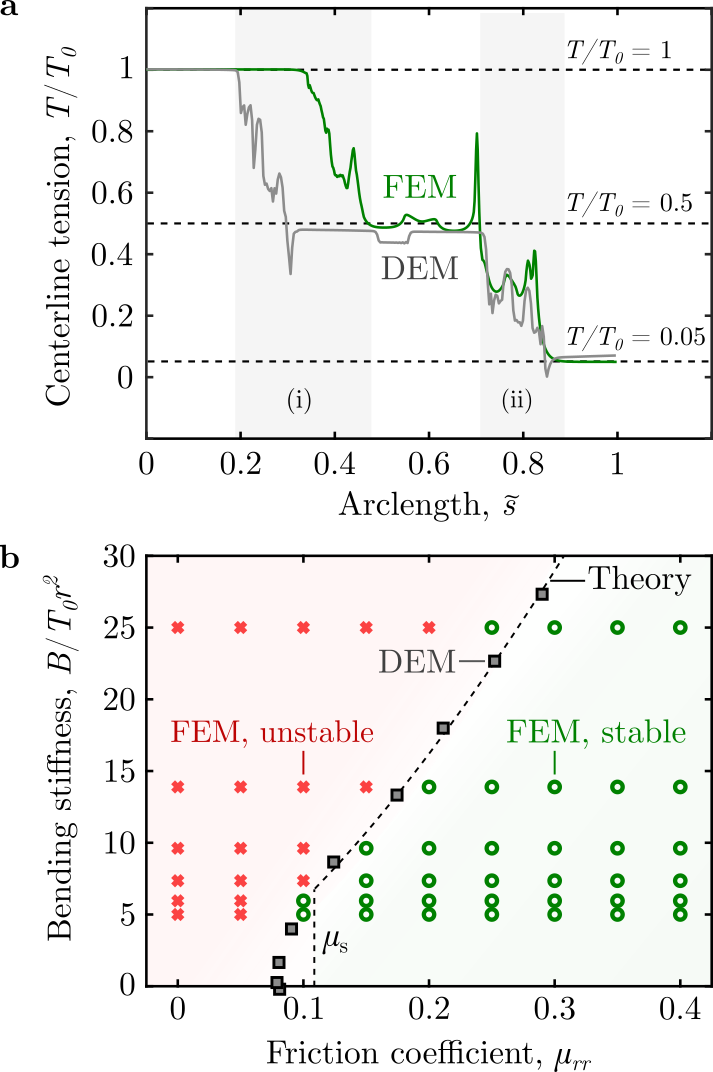}
    \caption{Tension profile and stability of the bowline. \textbf{a} Dimensionless tension profile versus arclength, $\tilde{s}{=}s/L$. Region (i) corresponds to the tension drop along the wrapping rod and (ii) in the central rod. \textbf{b} Stability diagram of the bowline in the parameter space of the normalized bending modulus of the rod versus the rod-rod friction coefficient. The knot is unstable (stable) in the red/left (green/right) regions, respectively. Crosses (circles) represent individual unstable (stable) FEM simulations, respectively. The stability boundaries predicted by DEM and our theory (fitted with $\alpha{=}1.69$, $\beta{=}10.6$; cf.~\eqref{eq:limits_2}) are represented by the squares and the dashed line, respectively.}
    \label{fig:fig4}
\end{figure}

Fig.~\ref{fig:fig4}\textbf{a} plots the tension profiles along the centerline under an applied load $T_{0}$, computed using both the FEM and DEM simulations, for $\mu_{\text{rr}}{=}0.37$. To focus on the mechanics of the locking unit, the cylinder is frictionless in the DEM simulations, and we set $\mu_{rc}{=}0.01$ in the FEM simulations (not zero, for numerical stability in FEM). A relatively small load, $T_{0}/20$, is applied at the free end, to also aid with numerical stability. In FEM, tension is obtained by summing nodal forces projected along the centerline direction across each cross-section. In DEM, this tension is directly available. We note that the centerline tension, $T(s)$, is normalized by $T_0{=}T(s{=}0)$ as $\tilde{T}(s){=}T(s)/T_0$. As shown in Fig.~\ref{fig:fig4}\textbf{a}, there are important tension variations along the centerline. First, there is a drop from $T_0$ to approximately $T_0/2$ in the loop formed by the center rod (at $0.2 \lesssim \tilde{s} \lesssim 0.5$; see Fig.~\ref{fig:fig4}\textbf{a},i). This first plateau value of $T_0/2$ is well explained by mechanical equilibrium: since the cylinder is nearly frictionless for this set of simulations, the tension at both extremities of the segment wrapping around the cylinder must be equal to half of the tension applied at the loading end. A second drop in tension from $T_0/2$ to approximately $0$ occurs around the wrapping rod region ($0.7 \lesssim \tilde{s} \lesssim 0.9$; see Fig.~\ref{fig:fig4}\textbf{a},ii). Importantly, both of these tension drops are directly linked to the locking unit highlighted in Fig.~\ref{fig:fig1}\textbf{d}. 

In addition to these global variations, we observe rapid tension fluctuations along the centerline that differ between FEM and DEM simulations. We attribute these differences to DEM's assumption of undeformable cross-sections, while cross-sections may deform in response to tension and curvature in flexible rods~\cite{Grandgeorge_Baek_Singh_Johanns_Sano_Flynn_Maddocks_Reis_2021, Grandgeorge_Sano_Reis_2022}. Additionally, the contact points and frictional force histories differ between FEM and DEM, potentially altering the initial configurations. However, given the bowline's reputed reliability, we do not expect these features to impact its overall stability.

\subsection{Reduced model for the bowline's stability}
\label{sec:elem_model}

The findings described above evidence that the bowline's locking performance primarily stems from a `locking unit' where one rod segment wraps around two core segments. We now analyze this critical region in more detail. The two observed tension drops appear essential to the bowline's stability, prompting us to develop a simple analytical model.

{\it Wrapping rod slippage condition.} The first tension drop identified in Fig.~\ref{fig:fig4}\textbf{a}(i) is due to the wrapping of a segment of the rod around two rods forming a central core (see Fig.~\ref{fig:fig1}\textbf{d}). This structure is similar to a one-turn capstan~\cite{euler1769remarques, eytelwein1832handbuch, Grandgeorge_Sano_Reis_2022}. For a perfectly flexible rod at the onset of sliding, the ratio of tensions at both ends is $T_{2}/T_{1}{=}e^{\mu_{\text{rr}}\varphi}$, with $\varphi$ the total angle swept by the rod; $\varphi{\approx}2\pi$ for the one-turn capstan considered here. With $T_0$ and $T_0/2$ being the tensions acting on both ends of the wrapping rod (see Fig.~\ref{fig:fig1}\textbf{e},i), to avoid sliding the rod-rod friction coefficient must satisfy the following condition:
\begin{equation}
    \mu_{\text{rr}} \ge \frac{\ln{2}}{\varphi} \coloneqq \mu_{\text{s}}.
    \label{eq:mus}
\end{equation} 
With $\varphi{\approx} 2\pi$, sliding is prevented if $\mu_{\text{s}} {\gtrsim} 0.11$. Since typical friction coefficients of physical ropes and filaments exceed this value of $\mu_{\text{s}}$, the bowline knot rarely experiences this failure mode in practice, provided that the knot is sufficiently tight to ensure full contact along the wrapping region.

{\it Central rods slippage condition.} The second tension drop (Fig.~\ref{fig:fig4}\textbf{a},ii) occurs along the central rods in the locking unit. The wrapping rod experiences an average tension $T{\simeq}(T_0+T_0/2)/2{=}3T_0/4$ (see Fig.~\ref{fig:fig1}\textbf{e}) that creates normal contact forces acting on the central rods. Approximating the wrapping rod as a perfectible flexible string (\textit{i.e.}, vanishing bending stiffness), the normal force $\text{d}N$ acting on a length $\text{d}s$ of contact is $\text{d}N{=}\kappa T~\text{d}s$, where $\kappa$ is the curvature of the wrapping rod. At the onset of sliding, the corresponding frictional force is $\text{d}F_{\text{s}}{=}\mu_{\text{rr}}\text{d}N$. Integrating the frictional force on the wrapping loop yields
\begin{equation}
    F_{\text{s}}=\int_{loop}\mu_{\text{rr}} \text{d}N = \mu_{\text{rr}} T \int_{loop} \kappa \text{d}s\approx2 \pi \mu_{\text{rr}} T.
    \label{eq:flexibleStrings}
\end{equation}
If we account for the finite bending stiffness of the wrapping rod, the normal force is no longer constant along contact~\cite{Baek_Johanns_Sano_Grandgeorge_Reis_2020}; thus, ~\eqref{eq:flexibleStrings} is not applicable and there is no simple theoretical prediction for $F_{\text{s}}$. The knot strength of the granny knot, whose locking unit exhibits a similar geometry to the bowline's, has been addressed recently~\cite{Johanns_Baek_Grandgeorge_Guerid_Chester_Reis_2023}. Combining experiments and FEM simulations, these authors showed that the knot strength scales nonlinearly as $\tilde T^\alpha$, where $\tilde T$ is the pre-tension of the knot and $\alpha{=} 1.56\pm 0.23$ is an experimentally determined, albeit robust, exponent. Following this study, we will assume that $F_{\text{s}} \propto \mu_{\text{rr}} T^\alpha$, where $\alpha$ is a to-be-determined exponent. The finite bending stiffness regime is expected to occur when tensions are small compared to the characteristic force $B \kappa^2$, where $\kappa$ is the centerline curvature of the wrapping rod. Combining this phenomenological description with \eqref{eq:flexibleStrings} yields:
\begin{subequations}  
\begin{align}
F_{\text{s}}&\approx2 \pi \mu_{\text{rr}} T \quad &\text { if }& T \gg  B\kappa^2,  \label{eq:Fs_vs_T_1}\\
F_{\text{s}}&\approx \beta \mu_{\text{rr}} T^\alpha (B \kappa^2)^{1-\alpha}\quad &\text { if }& T \ll  B\kappa^2. \label{eq:Fs_vs_T_2}
\end{align}
\end{subequations}  
The $(B\kappa^2)^{1-\alpha}$ term arises from dimensional analysis, and $\beta$ is a dimensionless constant.

Next, we consider both central rods to be a single, equivalent rod of radius $2r$. Since one of the central rods has almost no tension drop, and the other experiences a drop of approximately $T_0/2$ (see Fig.~\ref{fig:fig1}\textbf{e}), we take the effective tension drop in the equivalent rod to be $\Delta T=T_0/2$. For this equivalent rod to be secured by the wrapping rod, it must satisfy $\Delta T\le F_{\text{s}}$.  
Taking an average tension $T {\simeq} 3 T_0/4$ in the wrapping rod as described above, setting $\kappa{=}1/2 r$, and using Eqs.~(\ref{eq:Fs_vs_T_1},~\ref{eq:Fs_vs_T_2}), the stability condition $\Delta T\le F_{\text{s}}$ becomes:
\begin{subequations}  
\begin{align}
1 &\lesssim 3 \pi ~\mu_{\text{rr}} \quad &\text{if } T_0 \gg B/r^2, \label{eq:limits_1}\\ 
\frac{B}{T_{0}r^{2}}& \lesssim\left(\frac{3^{\alpha}\beta}{2}\right)^{\frac{1}{\alpha-1}}\mu_{\text{rr}}^{\frac{1}{\alpha-1}}\quad &\text{if } T_0 \ll B/r^2. \label{eq:limits_2}
\end{align}
\end{subequations}

For large tensions \eqref{eq:limits_1} gives $\mu_{\text{rr}} \ge 0.106$, which, coincidentally, is close to the wrapping rod slippage condition described above ($\mu_{\text{rr}} {\ge} \mu_{\text{s}} {\approx} 0.110$). For low applied tensions ($T_0 {\ll} B/r^2$),  \eqref{eq:limits_2} applies, and the value of $\mu_{\text{rr}}$ must increase beyond $\mu_{\text{s}}$ to prevent slippage of the central rods.

The simple model we just introduced postulates that the mechanics of the bowline's stability can be reduced to that of the simple \textit{locking unit} identified in Section~\ref{subsec:tension}. Specifically, our model predicts that the bowline can become unstable and unravel when either (i) the wrapping rod slips, or (ii) the two central rods slip. For both cases, this model offers an explicit stability boundary based on the rod's coefficient of friction, radius, and bending stiffness.

\subsection{Numerical simulations of the bowline's stability}

Next, to test the simple analytical model developed above, we conduct a systematic set of DEM and FEM simulations for different values of the normalized bending stiffness, $B/T_0 r^2$, and friction coefficient $\mu_{\text{rr}}$. 

FEM simulations are performed for each pair of $(\mu_{\text{rr}}, T_{0})$ values: $\mu_{\text{rr}}\in[0.0,\,0.4]$ with increments of 0.05 and $B/T_{0}r^2 \in \{5.0,\,6.0,\,7.4,\,9.1,\,13.9,\,25.0\}$, yielding a total of 54 simulations. First, a constant tension $T_{0}$ is applied at the loading end while fixing the free end, and the knot is allowed to reach static equilibrium. The free end is then released, and a small load $T_{0}/20$ is applied for numerical stability. The simulation runs until the knot either unravels because the applied load $T_{0}$ is too low (unstable) or remains tied, reaching equilibrium (stable). 

DEM simulations are performed as follows. A knot is prepared with $\mu_{\text{rr}}{=}0.35$ for a given value of $B/T_0 r^2$. A small load of $T_{0}/20$ is applied to the free end, similarly to FEM. Then, the friction coefficient is slowly decreased. At some critical value of $\mu_{\text{rr}}$, the knot becomes unstable and unravels. The two different modes of failure described in Section~\ref{sec:elem_model} --- slippage of the wrapping rod or the central rods  --- can then be identified with the measurements of the positions of the free ends and the locking unit at the onset of failure.

\subsection{Stability diagram of the bowline knot}

Fig.~\ref{fig:fig4}\textbf{b} presents a stability diagram in the $(\mu_{\text{rr}},\,B/T_0 r^2)$ parameter space, obtained from both FEM and DEM simulations. FEM simulations yield individual points in this parameter space that can be classified as unstable (red crosses) or stable (green circles). The boundary of stability as predicted by DEM is shown using black squares. 

First, we observe that DEM and FEM simulations agree well, with both predicting that the friction needs to increase to ensure stability when the applied tension is low compared to the characteristic force $B/r^2$. For very flexible rods ($B / T_{0} r^2 {\ll} 1 $), the stability occurs around $\mu_s {\approx} 0.09$, which agrees with the wrapping rod slippage mode in~\eqref{eq:mus}. For stiffer rods, we fit the stability boundary obtained from DEM simulations using~\eqref{eq:limits_2}.  This fitting procedure yields $\alpha_{\text{fit}} {=} 1.69 \pm 0.04$, which is remarkably compatible with the experimental value $\alpha{=}1.56 \pm 0.23$ measured  previously~\cite{Johanns_Baek_Grandgeorge_Guerid_Chester_Reis_2023}. For the prefactor of  the boundary line \eqref{eq:limits_2}, the fitting yields $\beta_{fit} {=} 10.6\pm 1.0$, in reasonable agreement with the expected crossover between the two functions in \eqref{eq:Fs_vs_T_1}--(\ref{eq:Fs_vs_T_2}) when $T{\sim} B\kappa^2$, which predicts $\beta \approx 2\pi$.

Despite the complex topology of the bowline, our simple stability model is able to capture, both qualitatively and quantitatively, the self-locking mechanics of the bowline. We note that the stability threshold given by Eq.~(\ref{eq:limits_2}) depends super-linearly ($1/(1-\alpha){\approx} 1.7$) on the friction coefficient. This high sensitivity aligns with the force-displacement results reported in Fig.~\ref{fig:fig3}\textbf{a}, where the pulling force after self-locking (at $\delta/r{\approx} 33$) doubles when $\mu_{\text{rr}}$ was increased by just 23\%.

\section{Conclusions}
\label{sec:conclusions}

Our results uncover a specific \textit{locking unit} in the bowline that plays a central role in determining the stability of this knot. Analysis of the tension distribution along the rod centerline revealed that most of the tension drop localizes in this region, where a single segment wraps around two others in a geometry reminiscent of a capstan. This observation motivated a simplified capstan-based stability model, which we validated against both finite-element (FEM) and discrete-element (DEM) simulations. The excellent agreement between theory, DEM, and FEM, even though DEM neglects cross-sectional deformations, suggests that simplified models can reliably capture the essential mechanics of tight knots.

More broadly, our study highlights mechanical motifs that extend beyond the bowline and recur across a variety of practical knots. Because similar locking units appear in sailing, climbing, and surgical applications, the present framework provides a foundation for identifying general stability criteria that couple topology, elasticity, and friction. While our focus has been on quasi-static behavior and simplified rod models, important open directions for future research include the role of dynamic effects, cyclic loading, and the multiscale structure of real ropes. We anticipate that these findings will encourage further studies of the bowline --- the `\textit{king of the knots}'--- as well as other designs whose performance critically depends on the interplay between geometry, elasticity, and friction.

\section*{Acknowledgements}

We are grateful to Paul Johanns for valuable discussions and support with preliminary experiments, and to Javier Sabater for assistance with the friction-coefficient measurements.

\appendix

\section{Centerline stretches using linear displacement mapping \label{app:directMapping}}

In this Appendix, we provide additional details on the mapping technique we introduced in Section~\ref{sec:mappingTechniqueMain} to set the initial configuration of the bowline knot in FEM simulations.

To avoid local high-curvature artifacts that may be introduced in the hand-drawn data, we first apply a uniform Laplacian smoothing on the discretized centerline coordinates. At the $n$-th smoothing iteration ($n{=}1,2, \dots, N$), the position of the $i$-th vertex along the deformed centerline, $\mathbf{x}_i$, is updated as
\begin{equation}
    \mathbf{x}_i^{n+1} = \mathbf{x}_i^{n} - \gamma \left[\mathbf{x}_i^{n}-\frac{1}{2}\left(\mathbf{x}_{i-1}^{n}+\mathbf{x}_{i+1}^{n}\right)\right],
    \label{eq:Laplacian_smoothing}
\end{equation}
where $\gamma > 0$ is a numerical factor that sets the strength of the smoothing. At each iteration stage, \eqref{eq:Laplacian_smoothing} re-positions each node closer to the average of its neighbors to smooth out local irregularities gradually; typically, we use $\gamma {=} 0.1$ and $N {=} 1000$, resulting in the smoothed coordinates $\mathbf{x}_i^N$ of a loose bowline knot. These coordinates are imported into MATLAB and scaled to the desired total arclength $L$ (we set $L{=}600$\,mm to match the experimental realization described in Section~\ref{sec:rod_fab}). Hereafter, we write $\mathbf{x}_i$ for the smoothed and scaled centerline coordinates.

\begin{figure}[h!]
    \centering
    \includegraphics[width=0.7\columnwidth]{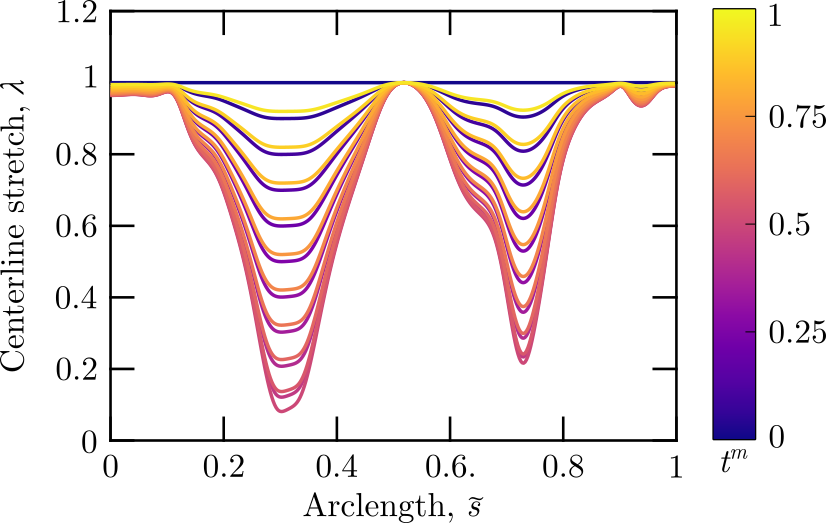}
    \caption{Centerline stretch, $\lambda$, as a function of the normalized rod arclength, $\tilde{s}$. Curves are color-coded as a function of mapping time, $t^{m}$. Linearly mapping centerline nodes to the desired knot shape results in compressive centerline strains of more than 80\% that impede convergence in FEM.}
    \label{fig:SuppA}
\end{figure}

Given the uniform mesh size $h$ along the centerline, we then obtain the desired displacement field, $\mathbf{u}_i = \mathbf{x}_{i}-\mathbf{X}_{i}$, for each centerline vertex. Here, $\mathbf{X}_{i}= ih\mathbf{e}_z$ denotes the centerline coordinates in the straight, undeformed configuration of the rod in the absolute Cartesian coordinate system with base vectors $(\mathbf{e}_{x},\mathbf{e}_{y},\mathbf{e}_{z})$. Note that it is generally not possible to directly impose the displacement $\mathbf{u}_i$ at all centerline nodes in FEM simulations. The reason is that, in practice, FEM solvers do not instantaneously apply the imposed displacement, but rather calculate the rod shape at a series of intermediate displacements between $\mathbf{0}$ and $\mathbf{u}_{i}$ to ensure convergence; these displacements correspond to intermediate shapes with centerline coordinates
\begin{equation}
    \mathbf{x}_{i}(t^{m})=\mathbf{X}_{i}+
    t^{m}\mathbf{u}_i,
\end{equation}
where the mapping iteration time $t^{m}$ varies in the range $t^{m}{\in}[0,\,1]$. As a simple, point-wise interpolation between the straight (undeformed) rod $\mathbf{X}_{i}$ and the desired knot shape $\mathbf{x}_{i}$, the intermediate shapes $\mathbf{x}_{i}(t^{m})$ tend to be characterized by extreme deformations of the centerline that impede convergence of the FEM solver (see~Fig.~\ref{fig:SuppA}). To circumvent this issue, we impose \textit{temporary displacements} $\mathbf{v}_i$, which are optimized to reduce stretching/compression of the centerline at intermediate rod shapes. The displacement mapping function we use is
\begin{equation} \label{eq:mapping}
    \mathbf{M}_i(t^{m}) =
    \mathbf{x}_{i}(t^{m})-\mathbf{X}_{i}=
    t^{m}\mathbf{u}_i + t^{m}(1-t^{m})\mathbf{v}_i.
\end{equation}
The factor of $t^{m}(1-t^{m})$ means that the temporary displacements first grow then decay in magnitude as $t^{m}$ varies over $[0,1]$; in particular, we recover the desired displacement, $\mathbf{u}_i$, at $t^{m}{=}1$. We choose the spatial part of the temporary displacements function, $\mathbf{v}_i$, to be a polynomial of degree $p$, and determine its optimal coefficients, $a_{j}^{\text{opt}}$ ($j{=}0,1, \dots, p$), by solving
\begin{equation}
     a_{j}^{\text{opt}} = \mathrm{argmin}_{a_j} \max_{t^{m}} \Lambda_{\text{max}}(a_{j}, t^{m}),
     \label{eq:param_opti}
\end{equation}
where $\Lambda_{\text{max}}(a_{j}, t^{m})$ is the maximum stretch function, defined as 
\begin{equation}
    \Lambda_{\text{max}}(a_{j}, t^{m}) = \max_{i}  \frac{||\mathbf{x}_{i+1}(t^{m})-\mathbf{x}_{i}(t^{m})||}{h},
\end{equation}
which returns the maximum stretch along the centerline at time $t^{m}$, and where $||\cdot||$ denotes the Euclidean norm.

The optimization problem specified in \eqref{eq:param_opti} is solved iteratively using a gradient descent algorithm. Once the maximum stretch reaches a chosen threshold (typically $|\Lambda_{\text{max}}-1|{<}0.3$), we use the optimized time- and node-dependent mapping function, $\mathbf{M}_i(t^{m})$, to automatically tie the knot in FEM using a custom \texttt{UDISP} user subroutine. Specifically, at each iteration time $t^{m}$, the algorithm performs:

\begin{algorithmic}\label{algo_disp}

\For{all nodes}
\If{node is on centerline and current node is $\mathbf{X}_i$} 
    \State $\mathbf{x}_{i}(t^{m}) \gets \mathbf{X}_{i}+\mathbf{M}_i(t^{m})$
\EndIf 
\EndFor

\end{algorithmic}

\section*{Suplementary Video Captions}

\noindent \textbf{Video 1:} A bowline is tied on a standard climbing rope around a horizontal, rigid cylinder. The knot is made taut by applying tension to the upper extremity of the rope while the other end remains free. The system remains stable after the large loop is severed using a cable cutter, provided that the locking unit is undisturbed. This demonstration highlights the importance of the locking unit in the stability of the bowline knot.\\ 

\noindent \textbf{Video 2:} 
Experimental procedure to tie a bowline on a VPS rod (see Fig. 1a and Section 4) around a horizontal rigid cylinder, prior to mechanical testing using a Universal Testing Machine (see Section 5). The following popular mnemonic is helpful for the tying process: `\textit{The rabbit comes out of the hole, runs around the tree, and goes back in the hole}.'

\bibliographystyle{elsarticle-num-names}
\bibliography{arXiv_main.bib}

\end{document}